\documentclass[twocolumn,osajnl,showpacs]{revtex4}

\usepackage{bm}
\usepackage{graphicx}

\begin{document}


\title{Effective fiber-coupling of entangled photons for quantum communication}

\author{Fabio~Antonio~Bovino}
\email{Fabio.Bovino@elsag.it} \affiliation{Elsag Spa., Via Puccini
2,  16154, Genova, Italy,}\homepage{http://www.elsag.it}

\author{Pietro~Varisco}
\email{Pietro.Varisco@elsag.it} \affiliation{Elsag Spa., Via
Puccini 2,  16154, Genova, Italy,}

\author{Anna~Maria~Colla}
\email{AnnaMaria.Colla@elsag.it} \affiliation{Elsag Spa., Via
Puccini 2,  16154, Genova, Italy,}

\author{Giuseppe~Castagnoli}
\email{Giuseppe.Castagnoli@elsag.it} \affiliation{Elsag Spa., Via
Puccini 2, 16154, Genova, Italy,}

\author{Giovanni~Di~Giuseppe}
\email{gdg@bu.edu} \altaffiliation[Also at~]{Istituto
Elettrotecnico Nazionale {\it G.~Ferraris}, Strada delle Cacce 91,
I-10153 Torino, Italy.} \affiliation{Quantum Imaging Laboratory,
Department of Electrical \& Computer Engineering, Boston
University, 8 Saint Mary's Street, Boston, Massachusetts 02215}

\author{Alexander~V.~Sergienko}
\homepage{http://people.bu.edu/alexserg/}\email{alexserg@bu.edu}
\affiliation{Quantum Imaging Laboratory, Department of Electrical
\& Computer Engineering and Physics, Boston University, 8 Saint
Mary's Street, Boston, Massachusetts 02215}
\date{\today}

\begin{abstract}
We report on theoretical and experimental demonstration of
high-efficiency coupling of two-photon entangled states produced
in the nonlinear process of spontaneous parametric down conversion
into a single-mode fiber. We determine constraints for the optimal
coupling parameters. This result is crucial for practical
implementation of quantum key distribution protocols with
entangled states.
\end{abstract}

\maketitle 

Entangled-photon pairs generated in the nonlinear process of
spontaneous parametric down conversion (SPDC) are proven to be a
highly desirable means~\cite{entangled} for practical quantum
cryptography~\cite{wal01-gis02}. The main difficulty of practical
utilization of such system usually stems from a relatively low
photon collection efficiency because of the complex spatial
distribution of SPDC radiation and due to the broad spectral width
of entangled-photon wave packets.

The problem of coupling entangled photons into a fiber has been
considered before by Kurtsiefer {\it et~al.}~\cite{kur01}.
Assuming the pump to be a plane wave the emission angle of the
SPDC has been calculated as a function of the wavelength.  The
waist of the focused pump beam has been chosen to maximally
overlap the "impression" of the Gaussian  mode of a single-mode
fiber on the crystal~\cite{kur01}. It has been pointed out that
the coupling efficiency may be significantly affected by
transverse walk-off.

In this letter we present a significantly modified approach
allowing us to achieve a high-efficiency coupling of the SPDC
pairs into single mode fibers. In particular, we demonstrate how
the pump beam waist, crystal length, optical system magnification
and the fiber mode field diameter (MFD) must obey a precise joint
relation in order to ensure high coupling efficiency. We describe
a specific model that allows us to determine a scaling law
accounting for all the real experimental parameters. We
demonstrate that obtained experimental data are in good agreement
with the proposed model.

The two-photons state can be written as~\cite{klyBOOK}
\begin{eqnarray}
   |\Psi\rangle_{{\rm SPDC}}=
     \int d\omega_{o}\,d{\bf q}_{o}
     \int d\omega_{e}\,d{\bf q}_{e}
     {\widetilde \Phi}({\bf q}_{o},\omega_{o};{\bf q}_{e},\omega_{e})
             \cdot\nonumber\\\cdot\,
     {\hat a}_{o}^{\dag}(\omega_{o},{\bf q}_{o})\,
     {\hat a}_{e}^{\dag}(\omega_{o},{\bf q}_{e})\,
   |0\rangle.
\end{eqnarray}
The function ${\widetilde \Phi}({\bf q}_{o},\omega_{o};{\bf
q}_{e},\omega_{e})=
         {\widetilde {\mathcal E}}_{p}({\bf q}_{o}+{\bf q}_{e},\,
                        \omega_{o}+\omega_{e})
         {\widetilde \chi}^{(2)}({\bf q}_{o},\omega_{o};{\bf q}_{e},\omega_{e})
$ accounts for the phase matching conditions. ${\widetilde
{\mathcal E}}_{p}(\cdot)$ represents the amplitude of the
plane-wave expansion of the pump field and ${\widetilde
\chi}^{(2)}({\bf q}_{o},\omega_{o};{\bf q}_{e},\omega_{e})=
     \int_{-L}^{0}dz\,
     \exp[-{\rm i}\,\Delta k({\bf q}_{o},\omega_{o};{\bf
     q}_{e},\omega_{e})\, z]$
with $\Delta k=k_{p}-k_{oz}-k_{ez}$. Inside the crystal the
$z$-component of the wave-vector is defined as $k_{z}({\bf
q},\omega)= \sqrt{[\omega\,n({\bf q},\omega)/c]^{2} -|{\bf
q}|^{2}}$. All the information on the state is given by the
amplitude ${\mathcal A}_{1,2}({\bf x}_{1},t_1;{\bf
x}_{2},t_{2})$~\cite{bah00,met01} of detecting the SPDC
two-photons in space-time events at $({\bf x}_{1},t_{1})$ and
$({\bf x}_{2},t_{2})$. The Fourier transform with respect to
$t_{1}$ and $t_{2}$ of the two-photon amplitude ${\mathcal
A}_{1,2}({\bf x}_{1},t_1;{\bf x}_{2},t_{2})$ is given by
\begin{eqnarray}\label{SPDC-Amplitude-1-freq}
    {\widetilde {\mathcal A}}_{1,2}({\bf x}_{1},\omega_o;{\bf x}_{2},\omega_e)
     =\int d{\bf q}_{o}\,d{\bf q}_{e}\,
         {\widetilde \Phi}({\bf q}_{o},\omega_{o};{\bf q}_{e},\omega_{e})
                 \cdot\nonumber\\\cdot
         {\mathcal H}_{1}({\bf x}_{1};{\bf q}_{o},\omega_{o})\,
         {\mathcal H}_{2}({\bf x}_{2};{\bf q}_{e},\omega_{e})\,
\end{eqnarray}
${\mathcal H}_{j}({\bf x}_{j};{\bf q},\omega)$ ($j=1,2$) being the
Fourier transform of impulse response functions $h_{j}({\bf
x}_{j},{\bf x};\omega)$ of the optical systems~\cite{bah00}
through which the two photons propagate from the output face of
the crystal to the detection plane.

The coupling of the photon pairs into fibers can be considered a
problem of maximizing the overlap between the two-photon amplitude
${\mathcal A}_{1,2}$ of entangled-photon state in the detector
plane with the field mode profiles of single-mode
fibers~\cite{kur01}. Assuming a {\it quasi-monochromatic} and {\it
quasi-plane} wave travelling in the $z$-direction and defined on
the two-dimensional continuous space of the detector planes, we
can express the electromagnetic field operator ${\hat
E}^{(+)}({\bf x},\omega)$ in terms of a linear superposition of
electromagnetic field operators, ${\hat c}_{lm}(\omega)$ and
${\hat c}_{k}(\omega)$ associated with a {\it complete
orthonormal} set of functions~\cite{blo90,gooBO}. Choosing
conveniently the noncontinuous set of {\it guided modes}
$\psi_{lm}({\bf x},\omega)$ and the continuous set of the {\it
radiation modes} $\psi_{k}({\bf x},\omega)$,
respectively~\cite{gooBO}, the field operator can be decomposed as
${\hat E}^{(+)}({\bf x},\omega)=
             \sum_{l,m}\,\psi_{lm}({\bf x},\omega)\,{\hat c}_{lm}(\omega)+
             \int dk\,\psi_{k}({\bf x},\omega)\,{\hat c}_{k}(\omega)$
where the new operators ${\hat c}_{\alpha}(\omega)$ are defined as
${\hat c}_{\alpha}(\omega)=
                 \int d{\bf x}\,
                 \psi^{\ast}_{\alpha}({\bf x},\omega)\,
                 {\hat E}^{(+)}({\bf x},\omega)$
with $\alpha=(l,m)$ or $k$, and obey the usual bosonic commutation
relations $[{\hat c}_{\alpha}(\omega),{\hat
c}^{\dag}_{\beta}(\omega)]=
                 \delta_{\alpha,\beta}$.
The amplitude in Eq.~(\ref{SPDC-Amplitude-1-freq}) can be expanded
in terms of guided and radiation modes. The coefficients of the
expansion for two guided modes, $(l,m)$ and $(l^{'},m^{'})$, are
given by
\begin{eqnarray}\label{tilde-A}
  {\widetilde {\mathcal A}}^{(1,2)}_{lm,l^{'}m^{'}}(\omega_{o},\omega_{e})
             = \int d{\bf x}_{1}\,d{\bf x}_{2}\,
  {\widetilde {\mathcal A}}_{1,2}({\bf x}_{1},\omega_o;{\bf x}_{2},\omega_e)
                 \cdot\nonumber\\\cdot
              \psi^{(1)\ast}_{lm}({\bf x}_{1},\omega_{o})\,
             \psi^{(2)\ast}_{l^{'}m^{'}}({\bf x}_{2},\omega_{e})
\end{eqnarray}

Coupling into a single-mode fiber can be quantified by the {\it
coupling efficiency} parameter $\eta_{{\rm fc}}$ defined as the
ratio of the probability to find two photons in the guided modes
over the square root of the product of the probability to find one
photon in a guided mode independently of the detection of the
other photon~\cite{joo94,mon98}. In the case of a single-mode
fiber, when only the {\it linearly-polarized} fundamental mode
${\it LP}_{01}$ $\psi^{(j)}_{01}$ is allowed~\cite{gooBO}, the
coupling efficiency takes the simple form
\begin{equation}\label{eta-fc-01}
   \eta_{{\rm fc}}=\frac{{\mathcal P}^{(1,2)}}
           {\sqrt{{\mathcal P}^{(1)}\cdot
                  {\mathcal  P}^{(2)}}}\,.
\end{equation}
The numerator of this expression is given by ${\mathcal
P}^{(1,2)}=\int d\omega_{o}\,d\omega_{e}| {\widetilde {\mathcal
A}}^{(1,2)}_{01,01}(\omega_{o},\omega_{e}) |^{2}$ where
${\widetilde {\mathcal A}}^{(1,2)}_{01,01}(\omega_{o},\omega_{e})$
is given by Eq.~(\ref{tilde-A}). The contributions at the
denominator of Eq.~(\ref{eta-fc-01}) are given by ${\mathcal
P}^{(1)}=
     \int d\omega_{o}\,d\omega_{e}
       \int d{\bf x}_{2}|\int d{\bf x}_{1}\,
  {\widetilde {\mathcal A}}_{1,2}({\bf x}_{1},\omega_o;{\bf x}_{2},\omega_e)\,
\psi^{(1)\ast}_{01}({\bf x}_{1},\omega_{o})|^{2}$ and analogous
expression for ${\mathcal P}^{(2)}$. The maximum coupling, i.e.
$\eta_{{\rm fc}}=1$, is reached when the two-photon amplitude in
Eq.~(\ref{SPDC-Amplitude-1-freq}) is the product of single-mode
field profiles  of two fibers.

We now consider a model  that includes propagation of both fields
through two equal infinite {\it ideal lenses} without an aperture
limit. We also assume that the output plane of the crystal, at a
distance $d_{{\rm bl}}$ from the lenses, is imaged on the fiber
plane, at a distance $d_{{\rm al}}$ from the lenses, i.e.
$1/d_{{\rm bl}}+1/d_{{\rm al}}=1/f$. The
amplitude~(\ref{SPDC-Amplitude-1-freq}) becomes ${\widetilde
{\mathcal A}}_{1,2}({\bf x}_{1},\omega_o;{\bf
x}_{2},\omega_e)\propto\Phi(\mu{\bf x}_{1},\omega_{o};\mu{\bf
x}_{2},\omega_{e})$ where $\mu=d_{{\rm bl}}/d_{{\rm al}}=d_{{\rm
bl}}/f-1$ is the inverse of the magnification and $\Phi({\bf
x}^{\prime},\omega_{o};{\bf x}^{\prime\prime},\omega_{e})$ is the
2-D inverse Fourier transform of the matching function
$\widetilde{\Phi}({\bf q}_{o},\omega_{o};{\bf q}_{e},\omega_{e})$,
which we calculate in the paraxial and quasi monochromatic
approximation. Inside the crystal, such approximations allow us to
consider only first terms in the exponential expansion of
expression ${\widetilde \chi}^{(2)}({\bf q}_{o},\omega_{o};{\bf
q}_{e},\omega_{e})$. For a type-II non-collinear configuration and
assuming a pump field to be factorable
  in terms of frequency and wave vectors ${\widetilde
{\mathcal E}}_{p}({\bf q}_{o}+{\bf
q}_{e},\,\omega_{p})={\widetilde {\mathcal
E}}_{p}(\omega_{p})\,{\widetilde {\rm E}}_{p}({\bf q}_{o}+{\bf
q}_{e})$, we can calculate the inverse Fourier transform with
respect to the frequencies and obtain $\Phi({\bf x}^{'},T+t/2;{\bf
x}^{''},T-t/2)\sim\Pi_{DL}(t)\,{\mathcal E}_{p}(T-\Lambda
t/D)\,{\rm E}_{p}[({\bf x}^{'}+{\bf x}^{''}-{\bf A}t/D)/2]\,
\delta({\bf x}^{'}-{\bf x}^{''}-{\bf B}t/D)$ where we have
introduced $t=t_{1}-t_{2}$ and $T=(t_{1}+t_{2})/2$. The function
$\Pi_{DL}(t)$ has value 1 for $0\leq t\leq DL$ and zero elsewhere.
The vectors ${\bf A}=2{\bf M}_{p}-{\bf M}$, ${\bf B}={\bf M}+2{\bf
Q}/{\bar{K}}$ depends on ${\bf M}$ and ${\bf M}_{p}$\,, which are
the spatial walk-off vectors for the extraordinary
fields~\cite{rub96} at the generation and pump frequency,
respectively. ${\bf Q}$ is the transverse wave-vector associated
with perfect phase matching along the intersection of cones, and
${\bar K}=2K_{o}K_{e}/(K_{o}+K_{e})$ represents a mean value of
wave-vector for generated photons inside the crystal. We have also
introduced $D=1/u_{o}-1/u_{e}$ and
$\Lambda=1/u_{p}-(1/2u_{o}+1/2u_{e})$ where $u_{j}$ is the group
velocity for the $j$-polarization. The expression has a simple
physical meaning. Due to the locality of the interaction, as
dictated by the delta function, photons are created in pairs at
each point of the crystal illuminated by the pump field, ${\rm
E}_{p}({\bf x})$. After their birth, photons propagate in the
dispersive nonlinear crystal environment experiencing longitudinal
$D$ and spatial walk-off, ${\bf M}$. They spread with respect to
each other in time and in transverse direction, according to the
travel distance, $z$. The transverse spread contributes through
two distinct processes when multiplied by the crystal length $L$.
The product $|{\bf A}|L$ represents the shift between the
generated pairs and the pump field, which is also extraordinary
and hence has a walk-off. $|{\bf B}|L$ is the spread between the
two entangled photons generated from the same pump photon. This
vector contains a contribution due to a spatial-walk-off, $|{\bf
M}|L$, and one more term , $2|{\bf Q}|\,L/{\bar{K}}$, which
represents the transverse distance between pairs generated at the
input face of the crystal with respect to the ones generated at
the output face, due to the geometry of optical propagation inside
the crystal. If we assume a pump beam with Gaussian profile, ${\rm
E}_{p}({\bf x})=\exp(-|{\bf x}|^{2}/2\,r^{2}_{p})$, as well as a
mode field profile, $\psi_{j}({\bf x})=\exp(-|{\bf
x}|^{2}/2\,w^{2})\,/\sqrt{\pi}w$, we can derive a closed
expression for the coupling efficiency, namely
\begin{eqnarray}\label{eta-c-model}
   \eta_{{\rm fc}}=4\frac{(1+\xi^{2})}
                 {(2+\xi^{2})^{2}}\,
                  \frac{{\rm erf}(\sigma_{c})}{\sigma_{c}}
                       \sqrt{\frac{\sigma_{1}}{{\rm erf}(\sigma_{1})}
                             \frac{\sigma_{2}}{{\rm erf}(\sigma_{2})}}
\end{eqnarray}
where $\xi=w\mu/r_{p}$ and
\begin{equation}
   \sigma_{c}=\frac{L}{r_{p}}
              \sqrt{\frac{(\alpha_{1}+\alpha_{2})\xi^{2}+\beta}
                         {\xi^{2}(2+\xi^{2})}}
                         \hspace{0.1cm};\hspace{0.2cm}
   \sigma_{j}=\frac{L}{r_{p}}
              \sqrt{\frac{\alpha_{j}}
                         {1+\xi^{2}}}
\end{equation}
The parameters $\alpha_{1}=|{\bf M}_{p}|^{2}+|{\bf Q}|^{2}/{\bar
K}^{2}$, $\alpha_{2}=|{\bf M}_{p}-{\bf M}|^{2}+|{\bf Q}|^{2}/{\bar
K}^{2}$, and $\beta=|{\bf M}|^{2}+4|{\bf Q}|^{2}/{\bar K}^{2}$ are
determined only by the crystal parameters and phase matching
geometry. The expression for $\eta_{{\rm fc}}$ depends on the size
of fibers effectively  imaged onto the crystal plane, $w\mu $, and
the crystal length, $L$, scaled to the pump beam waist into the
crystal, $r_{p}$.

\begin{figure}[ht]
   \centering
   \includegraphics[height=6cm, width=8cm]{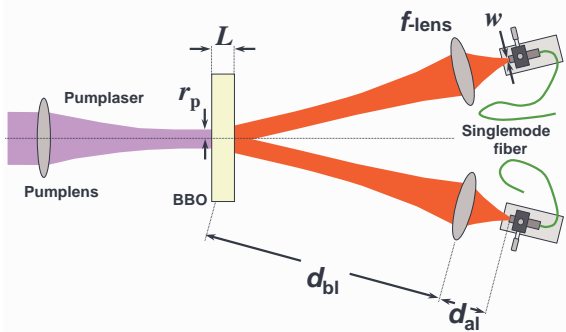}
   \caption{Sketch of the fiber coupling experiment.}\label{Figure-Setup}
\end{figure}

The experimental verification of the above model has been
performed using an actively mode-locked Ti:Sapphire laser, which
emitted pulses of light at 830~nm. After a second harmonic
generator, a 100-fsec pulse (FWHM) was produced at 415~nm, with a
repetition rate of 76~MHz and an average power of 200~mW. The
UV-pump radiation was focused to a beam diameter of 150~$\mu$m
inside a BBO crystal cut for a non-collinear type-II
phase-matching using a f=75~cm quartz lens (see
Figure~\ref{Figure-Setup}).

\begin{figure}[ht]
   \centering
   \includegraphics[height=5.5cm, width=7.5cm]{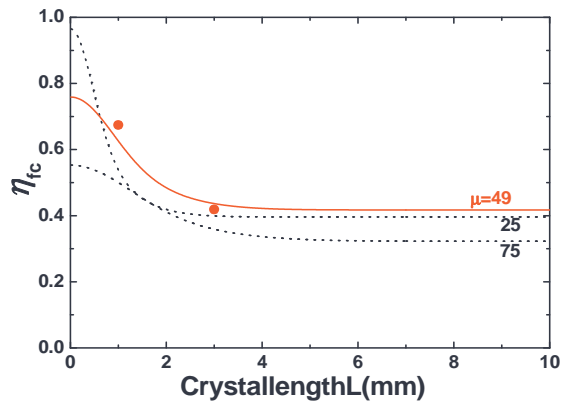}
   \caption{Coupling efficiency $\eta_{{\rm fc}}$ from
   Eq.~(\ref{eta-c-model}) vs. the crystal length $L$ for different
   values of parameter $\mu$. The solid line indicates the theoretical curve obtained
   using the parameters described in the text.
   The two points correspond to the experimental data.}\label{Figure-Fiber}
\end{figure}

Two points of  intersection of the ordinary and extraordinary
cones~\cite{kwi95} (3.5~deg relative to the pump direction) were
imaged into a single mode fiber with a MFD of 4.2~$\mu$m using two
f=15.4~mm coupling lenses placed 78~cm from the output face of the
non-linear crystal. We obtain a coupling efficiency of $\sim 18\%$
with a crystal of 3~mm length. We achieved  $\sim 29\%$ coupling
efficiency using a  1~mm crystal in the same experimental setup.
Taking into account the approximately 50\% detector efficiency and
85\% transmittance of filters, one can determine that effective
fiber coupling coefficients can reach 42\% and 68\%, respectively
(same results have been obtained with both 11~nm bandwidth filters
centered at 830~nm and pass-band filters). This result is
illustrated in Figure~\ref{Figure-Fiber}, where a solid line
indicates theoretical curve obtained from our model
Eq.~(\ref{eta-c-model}). The parameters used have been calculated
for crystal dispersion at 830~nm: $|{\bf M}_{p}|=0.07631$, $|{\bf
M}|=0.07243$, and $Q/{\bar K}=|{\bf M}|/2=0.036215$. The fiber
parameter is $w={\rm MFD}/2\sqrt{2}\sim1.48~\mu{\rm m}$, the pump
beam radius $r_{p}\sim$~53~$\mu$m, and $\mu\sim49$.

Examining the dependence of $\eta_{{\rm fc}}$ vs. crystal length
(see Figure~\ref{Figure-Fiber}) for different values of parameter
$\mu$ one can notice, for example, that one cannot reach a
coupling efficiency greater than $\sim50\%$ for
$r_{p}\sim$~53$~\mu$m and for crystals longer than 2~mm.

In conclusion, we have evaluated the dependence of coupling
efficiency of SPDC pairs into single-mode fiber on several major
experimental parameters. We obtained an analytical expression for
the coupling efficiency $\eta_{{\rm fc}}$ that allows us to
characterize the importance of the spatial walk-off and to choose
appropriate values of experimental parameters to reach high
coupling efficiency. Numerical inspection of the general
expression~(\ref{eta-fc-01}) with a more sophisticated model of
the impulse response function of our system can allow us to
further increase the coupling efficiency. {\it Acknowledgements}
This work was supported by MIUR (Project 67679). G.D.G and A.V.S.
also acknowledge support by DARPA and NSF.

\newpage



\end{document}